\documentclass[aps,pra,showpacs,twocolumn,groupedaddress,floatfix,longbibliography]{revtex4-1}

\usepackage{multirow}
\usepackage{graphicx}
\usepackage{amssymb}
\usepackage{pifont}
\usepackage{amsmath}
\usepackage{minibox}
\usepackage{enumerate}
\usepackage{mathtools}
\usepackage{wasysym}
\usepackage{color}
\usepackage{booktabs}

\begin{document}

\def\crta{\vrule height1.41ex depth-1.27ex width0.34em}
\def\dj{d\kern-0.36em\crta}
\def\Crta{\vrule height1ex depth-0.86ex width0.4em}
\def\Dj{D\kern-0.73em\Crta\kern0.33em}
\dimen0=\hsize \dimen1=\hsize \advance\dimen1 by 40pt

\title{How Secure are Two-Way Ping-Pong and LM05 QKD Protocols under a Man-in-the-Middle Attack?}

\author{Mladen Pavi{\v c}i{\'c}}

\email{mpavicic@irb.hr}

\affiliation{Center of Excellence for Advanced Materials and Sensors,
Research Unit Photonics and Quantum Optics, Institute Ruder
Bo\v skovi\'c, {10000 Zagreb}, Croatia;\\ 
Nanooptics, Department of Physics,
Humboldt-Universit{\"a}t zu Berlin, {12489 Berlin}, Germany}

\begin{abstract}
We consider a man-in-the-middle attack on two-way quantum
  key distribution ping-pong and LM05 protocols in which an
  eavesdropper copies all messages in the message mode, while being
  undetectable in the mode. Under the attack there is therefore no
  disturbance in the message mode and the mutual information between
  the sender and the receiver is always constant and equal to one and
  messages copied by the eavesdropper are always genuine. An attack
  can only be detected in the control mode but the level of detection
  at which the protocol should be aborted is not defined. We examine
  steps of the protocol to evaluate its security and find that the
  protocol should be redesigned. We also compare it with the security
  of a one-way asymmetric BB84-like protocol in which one basis
  serves as the message mode and the other as the control mode but
  which does have the level of detection at which the protocol should
  be aborted defined.
\end{abstract}



\maketitle

\section{Introduction}
\label{intro}

Quantum cryptography, in particular quantum key distribution 
(QKD) protocols, offers us, in contrast to the classical one,
provably unbreakable communication based on the 
quantum physical properties of the information carriers 
\cite{elliot-DARPA,sasaki-tokyo-qkd-10,peev-zeilinger-njp09}.
So far, implementations were mostly based on the one-way BB84 
protocol~\cite{bb84} which is unconditionally secure provided the 
quantum bit error rate (QBER) is low enough. However, 
QBER in BB84-like protocols might be high and since we cannot 
discriminate eavesdropper's (Eve's) bit flips from bit flips 
caused by noise in the line, the request of having QBER low 
enough for processing the bits is often difficult to satisfy.  
E.g., 4-state BB84 with more than 0.11 \cite{scarani-09} 
and 6-state BB84 \cite{bruss} with more than 0.126 \cite{scarani-09}
disturbance ($D$) have to be aborted ($D$ is defined as the
amount of polarization-flips caused by Eve, maximum being 0.5).
$D$ includes the inherent QBER as well as possible Eve in the line.
If Eve were the only cause of $D$, the mutual information
between the sender (Alice) and Eve ($I_{AE}$) wound surpass the one
between Alice and the receiver (Bob) ($I_{AB}$): $I_{AE}>I_{AB}$ for
$D>0.11, 0.126$, respectively. 

Protocols using two-way quantum communications have also been
proposed. Since they are less efficient versions of BB84 protocols,
they have no meaningful advantage. Here we show that the security
of some two-way protocols is vulnerable under a man-in-the-middle
(MITM) attack. In particular, entangled photon two-way protocols
based on two \cite{bostrom-felbinger-02} (also called a
{\em ping-pong\/} (pp) protocol) and four ($\Psi^\mp,\Phi^\mp$)
\cite{cai-li-04} Bell states, on the one hand and a single photon
deterministic Lucamarini-Mancini (LM05) protocol, on the other
\cite{lucamarini-05,lucamarini-mancini-13}. Several varieties, 
modifications, and generalisations of the latter protocol are given
in \cite{henao-15,khir-12,shaari-mancini-15,pirandola-nat-08}.
Two varieties were implemented in \cite{cere-06} and 
\cite{kumar-lucamarini-08}. The former pp protocol was implemented
by Ostermeyer and Walenta in 2008~\cite{ostermeyer-08} while the 
protocol with four Bell states cannot be implemented with linear
optics elements \cite{luetkenhaus-99,vaidman99}. In the
aforementioned references various security estimations have been
obtained. 

In \cite{lu-cai-11} Lu, Fung, Ma, and Cai provide a security proof
of an LM05 deterministic QKD for the kind of attack proposed in
\cite{lucamarini-05,lucamarini-mancini-13}. Nevertheless,
they claim it to be a proof of the unconditional security of LM05.
In \cite{han-14} Han, Yin, Li, Chen, Wang, Guo, and Han provide
a security proof for a modified pp protocol and prove its security
against collective attacks in noisy and lossy channel.

All previous elaborations of the security of two-way protocols 
assume that Eve attacks each signal twice, once on the way from
Bob to Alice, and later on its way back from 
Alice to Bob, and that in doing so she disturbs the signal in the
message mode.

However, there is another attack in which an undetectable Eve
encodes Bob's signals by mimicking Alice's encoding of a decoy
signal sent to her which we elaborate on in this paper. We
consider the two-way deterministic QKD protocols under a
MITM attack where, Eve---undetectable in the
message mode (MM)---can acquire all the messages, meaning that
there is no disturbance in the MM ($D_{\rm MM}$) at all. In the
control mode (CM) there is a disturbance ($D_{\rm CM}$), 
but there is no critical $D$ at which Alice and Bob should abort
the protocol. The only way to delete bits of the raw key snatched
by Eve is via privacy amplification and for disturbances close to
$D_{\rm CM}=0.5$, when Eve is in the line all time, it seems
impossible to distinguish whether Eve has or has not obtained the
whole key. In order to verify that conjecture, we prove that
the security proof carried in \cite{lu-cai-11} does not cover a
MITM attack and that therefore cannot be called ``unconditional.''

We also compare two-way protocols under a MITM attack with a
recent one-way asymmetric BB84-like protocol \cite{bunandar-18}
in which the $\{|0\rangle,|1\rangle\}$ basis serves as MM and
the $\{|\pm\rangle\}$ basis as CM, under a MITM attack. The latter
protocol resolves the problem of absence of
inherent critical disturbance by introducing a predetermined
threshold disturbance after which Alice and Bob abort the protocol.
This makes the protocol conditionally secure and we propose a
similar solution for the two-way protocols.

In Sec.~\ref{sec:entangled} we present the protocols and MITM
attacks on them. In Sec.~\ref{sec:security} we discuss the security
of two-way protocols and analyze their proof of unconditional security;
we also compare properties of two-way protocols with those of the
standard BB84 and the aforementioned asymmetrical BB84-like one.
In Sec.~\ref{sec:conclusion} we present some concluding points and
a summary of the results achieved in the paper.

\bigskip
\section{Protocols and Attacks on Them}
\label{sec:entangled}

Ping-pong (pp) protocol is based on two Bell states
\cite{bostrom-felbinger-02}. Bob prepares entangled photons in one of
the Bell states, sends one of the photons to Alice while keeping the
other one in a quantum memory (qm). Alice either returns the photon as
is or acts on it so as to put both photons into another Bell states.
Bob combines the photon he receives from Alice with the one he kept in
qm at a beam splitter (BS) to decode Alice's messages. The messages
are said to be sent in {\em message mode\/} (MM). There is also a
{\em control mode\/} (CM) in which Alice measures Bob's photon and
announces her outcomes over a public channel. 

Bell basis used in the pp protocol consists of two Bell states
\begin{eqnarray}
|\Psi^\mp\rangle=\frac{1}{\sqrt{2}}(|H\rangle_1|V
\rangle_2\mp|V\rangle_1|H\rangle_2),
\label{eq:bell-states} 
\end{eqnarray}
where $|H\rangle_i$ ($|V\rangle_i$), $i=1,2$, represent horizontal
(vertical) polarized photon states.

Photon pairs in the state $|\Psi^-\rangle$ are generated by a
down-converted entangled photon source. To send $|\Psi^-\rangle$
state Alice just returns her photon to Bob. To send $|\Psi^+\rangle$
she puts a half-wave plate (${\rm HWP}(0^\circ)$) in the path of her
photon. The HWP changes the sign of the vertical polarization. 

At Bob's BS the photons in state $|\Psi^-\rangle$ will split and
those in state $|\Psi^+\rangle$ will bunch together. 

\begin{figure*}[ht]
\center
\includegraphics[width=0.99\textwidth]{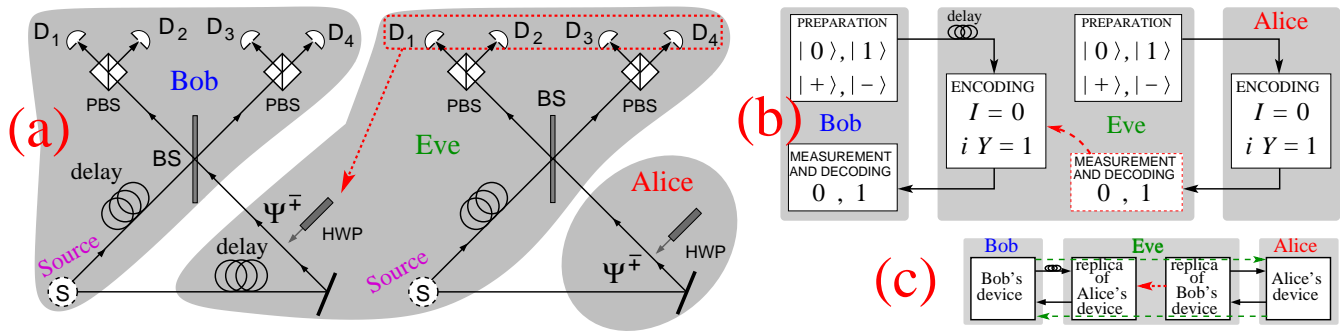}
\caption{(a) Nguyen's attack \cite{nguyen-04} by which Eve is able
  to deterministically (and undetectably in the MM) copy every one
  of the Bell-state messages in the pp protocol
  \cite{bostrom-felbinger-02};
  (b) Lucamarini's attack \cite[p.~61, Fig.~5.5]{lucamarini-phd-03}
  by which Eve is able to deterministically (and undetectably in the
  MM) copy every message in the LM05 protocol; (c) Common schematics
  of both attacks; The green dashed line shows the path of photons
  when Eve is not in the line.}
\label{fig:2w}
\end{figure*}

Eavesdropper Eve carries out a MITM, 
designed by Nguyen \cite{nguyen-04}, as follows. She puts Bob's
photon in a qm (delays it in fiber coil) and makes use of a copy
of Bob's device to send Alice a photon from a down-converted pair
in state $|\Psi^-\rangle$ as shown in Fig.~\ref{fig:2w}(a).
When Eve receives the photon from Alice she combines it with the
other photon from the pair and determines the Bell state in the
same way Bob would. She uses this result to generate the same Bell
state for Bob by putting the appropriate HWPs in the path of Bob's
photon. 

Thus, Eve is able to copy every single message in the MM
undetectably and therefore sending messages in the MM is
equivalent to sending plain text ``secured'' by the CM.

In the LM05 protocol \cite{lucamarini-phd-03,lucamarini-05} Bob
prepares a qubit in one of the four states
$|0\rangle$,  $|1\rangle$ (the Pauli $\boldsymbol Z$ eigenstates),
$|+\rangle$, or $|-\rangle$ (Pauli $\boldsymbol X$ eigenstates) and
sends it to his counterpart Alice. In the MM she modifies the qubit
state by applying either $\boldsymbol I$, which leaves the qubit
unchanged and encodes the logical {\bf 0}, or by applying
$i{\boldsymbol Y}={\boldsymbol Z}{\boldsymbol X}$, which flips 
the qubit state and encodes the logical {\bf 1}.
($i{\boldsymbol Y}|0\rangle=-|1\rangle$, 
$i{\boldsymbol Y}|1\rangle=|0\rangle$, 
$i{\boldsymbol Y}|+\rangle=|-\rangle$, 
$i{\boldsymbol Y}|-\rangle=-|+\rangle$.) 
Alice now sends the qubit back to Bob who measures it in the same
basis in which he prepared it and deterministically infers Alice’s
operations, i.e., her messages, without basis reconciliation procedure.

Eavesdropper Eve carries out a MITM, 
designed by Lucamarini \cite[p.~61, Fig.~5.5]{lucamarini-phd-03}, as
follows. As shown in Fig.~\ref{fig:2w}(b). Eve delays Bob's photon
(qubit) in a fiber spool (a quantum memory) and sends her own decoy
photon in one of the four states $|0\rangle$, $|1\rangle$, $|+\rangle$,
or $|-\rangle$ to Alice, instead. Alice encodes her message via
$\boldsymbol I$ or $i{\boldsymbol Y}$ and sends the photon back.
Eve measures it in the same basis in which she prepared it, reads
off the message, encodes Bob's delayed photon via ${\boldsymbol I}$,
if she read {\bf 0}, or via $i{\boldsymbol Y}$, if she read {\bf 1},
and sends it back to Bob. 

Eve never learns the states in which  Bob sent his photons but
that is irrelevant in MM since only polarization flipping
or not flipping encode messages. Alice also need not know Bob's
states \cite{lucamarini-05}. Eve could only be revealed in
CM in which Alice carries out a projective measurement of the
qubit along a basis randomly chosen between ${\boldsymbol Z}$
and ${\boldsymbol X}$, prepares a new qubit in the same state as
the outcome of the measurement, sends it back to Bob, and reveals
this over a classical public channel
\cite{lucamarini-05}.

\bigskip
\section{Security of the Protocols}
\label{sec:security}

To reach the main point of the paper, let us first discuss the
one-way asymmetric (aBB84) and symmetric (sBB84, i.e., standard
BB84) protocols. 

A recent definition of aBB84 \cite{toma-gisin-renner-12} reads:
``Alice [asks her] entanglement-based source to [randomly] prepare
quantum states in two bases, ${\mathbb X}=\{|0\rangle,|1\rangle\}$
and ${\mathbb Z}=\{(|0\rangle+|1\rangle)/\sqrt{2},
(|0\rangle-|1\rangle)/\sqrt{2}\}$\dots\ Bob [randomly]
measure[s] quantum systems in [these two] bases\dots\ The protocol
is {\em asymmetric} [meaning that] the number of bits measured in
the two bases ($n$ bits in the ${\mathbb X}$ basis and $k$ bits in
the ${\mathbb Z}$ basis) are not necessarily equal
\cite{lo-chau-ard-05}\dots\ {\bf Sifting:} Alice and Bob
broadcast their basis choices over the classical
channel\dots\ {\bf Error correction:} (EC) A reconciliation
scheme that broadcasts [chosen] bits of classical error correction
data is applied. Bob compute[s] an estimate $\hat{\mathbf Y}$ of
the raw key string $\mathbf Y$. Alice computes universal$_2$ hash
function of $\mathbf Y$ [and] sends [it] to Bob. If the hash[es]
of $\hat{\mathbf Y}$ and $\mathbf Y$ disagree, the protocol aborts.
{\bf Privacy amplification:} (PA) Alice extracts $l$ bits of secret
key $\mathbf S$ from $\mathbf Y$ using a random universal$_2$ hash
function. The choice of function is communicated to Bob, who uses
it to calculate $\mathbf S$.'' \cite[p.~3]{toma-gisin-renner-12}
There are other similar definitions of aBB84 in the literature
\cite{lo-chau-ard-05,scarani-renner-08,cai-scarani-09,zhou-14,mizutani-15}. 

When $n=k$, aBB84 turns into sBB84, i.e., it becomes
identical to the original BB84. In what follows, when not
explicitly stated otherwise, under BB84 we mean sBB84. 

What is essential for the standard aBB84 and sBB84, is that Eve
cannot avoid introducing disturbance ($D$).
Specifically, when Alice sends messages in ${\mathbb X}$
and ${\mathbb Z}$ bases, Eve cannot avoid introducing $D$.
E.g., Alice sends $|1\rangle$ in ${\mathbb X}$ basis,
Eve reads it as $(|0\rangle+|1\rangle)/\sqrt{2}$ in
${\mathbb Z}$ basis and resends it to Bob who, say in ${\mathbb X}$
basis, reads it either as $|0\rangle$ or as $|1\rangle$.
If the former, it will be discarded in the EC procedure, if the
latter, it will be accepted as a valid message. As for Eve, no
public information can enable her to find out what Bob actually
measured hence she loses information.

\begin{figure*}[ht]
  \begin{center}
    \includegraphics[width=0.325\textwidth]{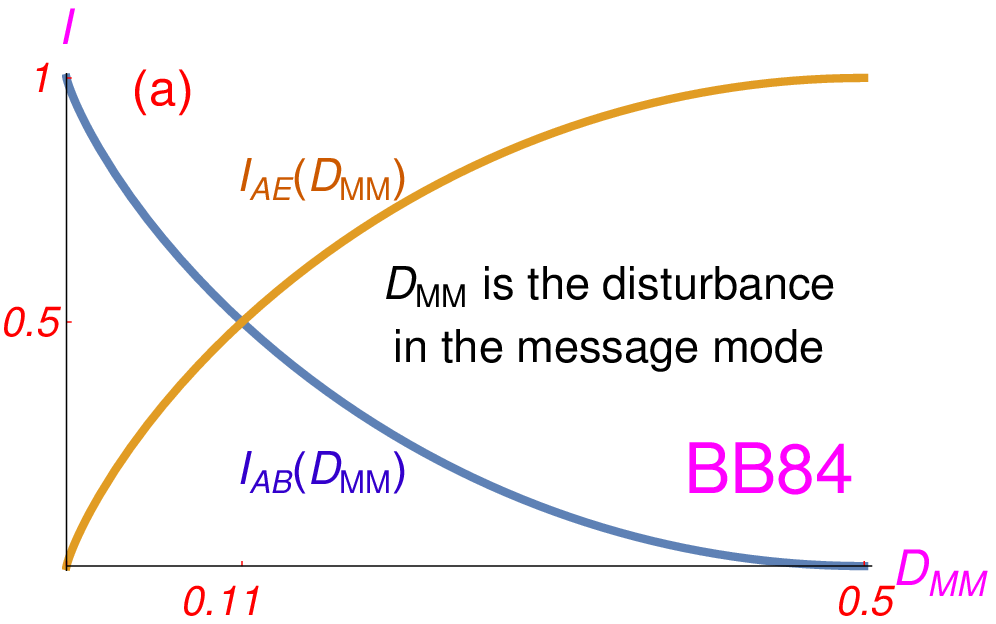}
    \includegraphics[width=0.325\textwidth]{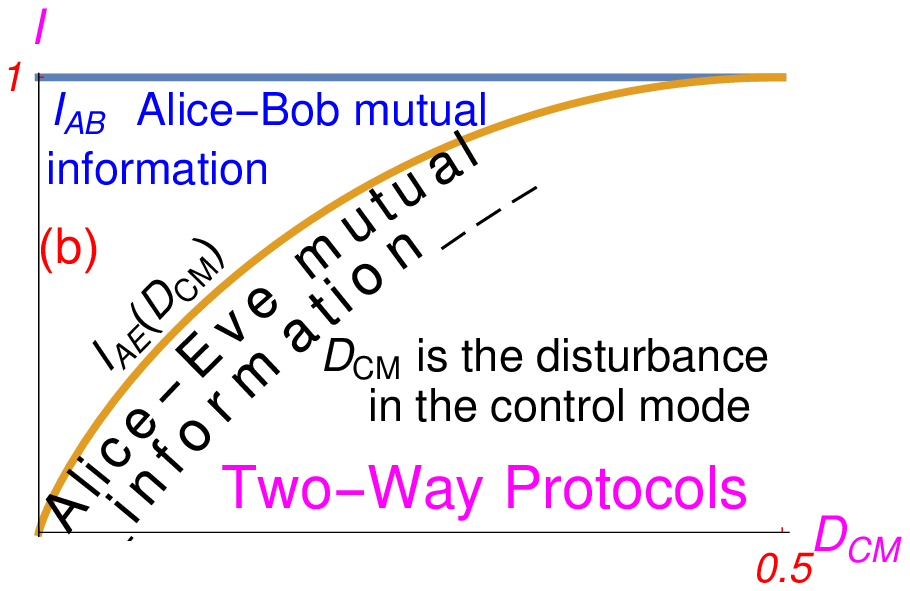}
    \includegraphics[width=0.325\textwidth]{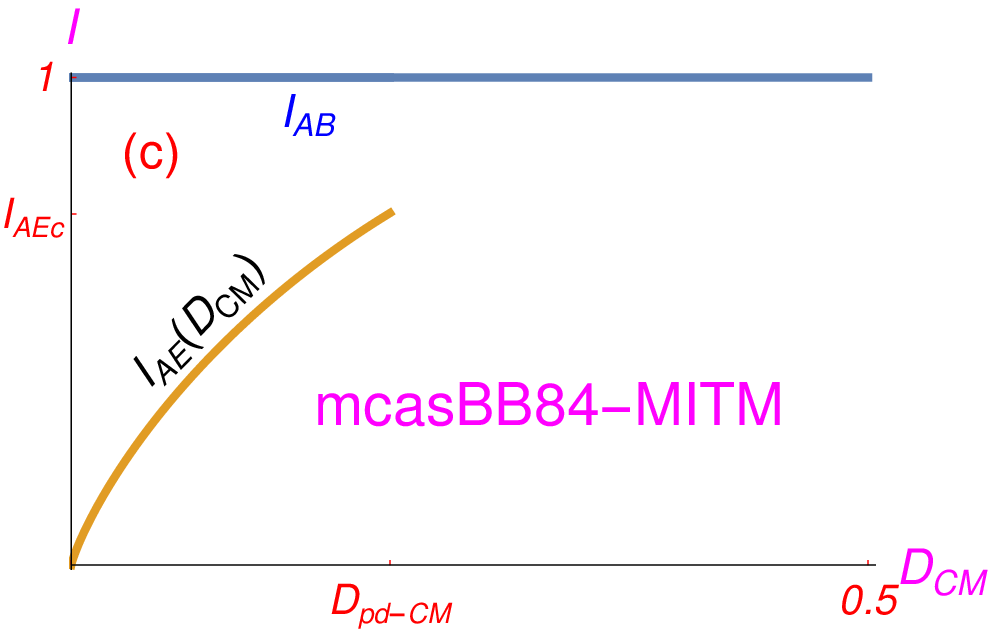}
  \end{center}
  \caption{Mutual information plots for (a) one-way protocol BB84;
    (b) two-way protocols with either pp entangled Bell states or
    with LM05-like single photon states under a MITM attack;
    (c) one-way asymmetric BB84-like protocol, in which one basis
    serves as MM and the other as CM, under a MITM attack
    (mcasBB84-MITM); $I_{AEc}$ stands for $I_{AE}(D_{pd\rm -CM})$.}
  \label{fig:dist}
  \end{figure*}

However, Alice and Bob lose their information too, and 
as shown in Fig.~\ref{fig:dist}(a) in a BB84 protocol, when the
level of disturbance approaches $D_{\rm MM}=0.11$ (the MM is the
only mode of the standard BB84 protocol) the mutual
information between Alice and Eve $I_{AE}$ approaches the mutual
information between Alice and Bob $I_{AB}$ and they have to abort the
protocol. Note that $I_{AB}=1+D\log_2D+(1-D)\log_2(1-D)$
and $I_{AE}=-D\log_2D-(1-D)\log_2(1-D)$
\cite{fuchs-gisin-peres-97} and that, ideally, for $D_{\rm MM}<0.11$,
EC can eliminate all errors induced by Eve and that PA can remove
all key bits Eve might have collected, no matter how close to 0.11
$D_{\rm MM}$ is. This is so because both $I_{AB}$ and $I_{AE}$ are
functions of $D_{\rm MM}$, i.e., functions of the disturbance in
the message mode for which the mutual information in the very same
message mode is calculated. The closer $D_{\rm MM}$ is to 0.11, the
more difficult is for Alice and Bob, after PA, to extract the secure
key from the raw key, since the former becomes smaller and smaller.
``The efficiency of privacy amplification rapidly decreases when
[$D_{\rm MM}$] increases'' \cite[p.~165, mid right column]{gisin-02}.
``At $D_{\rm MM}=0.11$ the secure-key length formally vanishes''
\cite[p.~524]{molotkov-07}. See also \cite{fuchs-gisin-peres-97}.

For a MITM attack on two-way protocols (which are without any
sifting), when Eve is in the line all the time, there is no
$D_{\rm MM}$ that Eve induces and the mutual information between
Alice and Bob as well as between Alice and Eve is equal to
unity: $I_{AB}=1$. In Fig.~\ref{fig:dist}(b) $D_{\rm CM}$ indicates
the presence of Eve, where $D_{\rm CM}=0.5$ would mean that Eve is
always in the line. For $D_{\rm CM}$ slightly below $D_{\rm CM}=0.5$
one cannot exclude the possibility that Eve has all the messages.

When Eve has all the messages then there is no Alice-Bob ``privacy''
they could amplify. When Eve snatches only a portion of messages,
then, for $D_{\rm CM}$ close to 0.5, there is still a question
whether Eve has all messages or not and whether Alice and Bob can
erase Eve's messages with their PA\@. With that in mind, we can
examine the security evaluation for the MITM and verify whether
the proofs of unconditional security carried out for other kind
of attack on LM05 in \cite{lu-cai-11,lucamarini-mancini-13} might
apply to it as well. 

In the aforementioned security proof \cite{lu-cai-11}, which is
claimed to be {\em unconditional}, the authors assume that Eve
probes the qubit, entangling it with an ancilla. However, their
approach does not cover the MITM attacks\@. To show this we point
to the following steps in their proof of unconditional security
vs.~two-way-protocol-under-MITM counter-steps:
\begin{itemize}
\item[{\bf \cite{lu-cai-11}}] p.~2, 2nd paragraph from the top:
  ``Alice announces partial of her key bits in the encoding
  mode [MM]. They compute
  the error rate $e$ in the Alice-Bob channel.''
\item[\color{red}{\textsf{MITM}}] {\em $I_{AB}=1$ and Eve does
    not induce any error in the {\rm MM}, ever.}
\item[{\bf \cite{lu-cai-11}}] p.~2, Sec.~III.A: ``Eve cannot
  gain any information about Alice’s key bits if she only
  attacks the qubits after Alice’s encoding operation.''
\item[\color{red}{\textsf{MITM}}] {\em Since Eve in her MITM sends
    her own photons to Alice and then reads off $\boldsymbol I$ or
    $i{\boldsymbol Y}$ from Alice's encoding of those qubits, Eve
    gains all information from Alice's qubits, more precisely, from
    Eve's qubits encoded by Alice. Note the neither Alice nor Eve
    know which states the qubits Bob sends are in. They only control
    $\boldsymbol I$ and $i{\boldsymbol Y}$.}
\item[{\bf \cite{lu-cai-11}}] Eve’s most general quantum operation
  can be described by a unitary operation together with an ancilla.
  In the Bob-Alice channel, when Bob sends a qubit in state
  $|0\rangle$ and Alice measures in the basis $|0\rangle$,$|1\rangle$,
  she will get the measurement outcomes $|0\rangle$ with probability
  $c_{00}^2$ or $|1\rangle$ with probability $c_{01}^2$.
\item[\color{red}{\textsf{MITM}}] {\em Alice does not measure qubits.
    She just applies $\boldsymbol I$ and $i{\boldsymbol Y}$.}
\item[{\bf \cite{lu-cai-11}}] Eve’s most general attack (with
  ancillas) is\hfill\break
  $U_{BE}|0\rangle_B|E\rangle=c_{00}|0\rangle_B|E_{00}\rangle+
     c_{01}|1\rangle_B|E_{01}\rangle$,
     
$U_{BE}|1\rangle_B|E\rangle=c_{11}|1\rangle_B|E_{11}\rangle+
  c_{10}|0\rangle_B|E_{10}\rangle,$

$U_{BE}|+\rangle_B|E\rangle\!=\!c_{++}|+\rangle_B|E_{++}\rangle\!+\!
     c_{+-}|-\rangle_B|E_{+-}\rangle$,

$U_{BE}|-\rangle_B|E\rangle\!=\!c_{--}|-\rangle_B|E_{-+}\rangle\!+\!
     c_{-+}|+\rangle_B|E_{-+}\rangle.$
     
Fidelities are $f_0=c_{00}^2$, $f_1=c_{11}^2$, $f_+=c_{++}^2$,
and $f_-=c_{--}^2$. $f_0=f_1$ and $f_-=f_+$ are assumed\dots

Bob’s qubit is in a mixed state
$\rho^B=(|0\rangle\langle 0|+|1\rangle\langle 1|)/2$.
The joint state of the forward qubit and Eve’s ancilla becomes
$\rho_{BA}^{BE}=U_{BE}(\rho^B\otimes|E\rangle\langle E|)U_{BE}$.
Alice's encoded qubit together with Eve’s ancillas is: 
$\rho^{ABE}=\frac{1}{2}|0\rangle\langle 0|^A\otimes \rho^{BE}_{BA}
+ \frac{1}{2}|1\rangle\langle 1|^A\otimes
{\boldsymbol Y}_B\rho^{BE}_{BA}
{\boldsymbol Y}_B$\dots

The asymptotic key generation rate is
$r=\lim_{m\to\infty}\frac{k(m)}{m}$, where $m$ is the size of
the raw key and $k(m)$ is the number of the final key bits.
Alice sends Bob EC information over a classical channel so
that he can correct his raw key to match Alice’s.
\item[\color{red}{\textsf{MITM}}] {\em Eve does not induce any
    disturbance, so there is no EC.} 
\item[{\bf \cite{lu-cai-11}}] The final key is then derived by
  applying two-universal hashing to their common raw key as PA\@.
  The secure key rate $r_{\rm PA}$ for secret key generation is
  bounded by the conditional entropy of Alice and Bob’s key
  bits given the quantum information of Eve about the key bits
$r_{\rm PA}=S(\rho^A|\rho^{BE})=
-{\rm }tr\rho^{ABE}\log_2\rho^{ABE}+{\rm tr}\rho^{BE}\log_1\rho^{BE}=
1-h(\xi)$, where $\xi=c_{++}^2-c_1^2$, $c_1=c_{01}=c_{10}$,
and $h(x)=-x\log_2x-(1-x)\log_2(1-x)$
is the binary Shannon entropy. In particular, if Eve does not
attack the forward qubits in the Bob-Alice channel, i.e.,
$f_0=f_1=f_+=f_- = 1$, one can find that
$r_{\rm PA}(\xi)=1$. This states that Eve cannot gain
any information about Alice’s key bits if she does not attack
the travel qubit in the Bob-Alice channel first. Consider the
case that Eve measures each forward qubit in the Bob-Alice
channel in the basis $|0\rangle,|1\rangle$. Alice and Bob can
verify that $f_0=f_1=1$, and $f_+=f_-=\frac{1}{2}$. In this case,
we have $r_{\rm PA}(\xi)=0$. On the other hand, Eve can also measure
each forward qubit in the Bob-Alice channel in the basis
$|+\rangle,|-\rangle$, which gives $f_+=f_-=1$ and
$f_0=f_1= \frac{1}{2}$, and thus $r_{\rm PA}(\xi)=0$. That is, Eve can
gain full information of Alice’s key bits if she has exactly known
the forward states before Alice’s encoding operations.
\item[\color{red}{\textsf{MITM}}] {\em Eve does not measure qubits
    (or ancillas). When she is in the line all the time, she just
    reads off $\boldsymbol I$ and $i{\boldsymbol Y}$ Alice executed
    on her qubits and applies them to Bob's qubits she stored, i.e.,
    she copies the whole key---both sides have the whole key.
    There is no privacy which can be amplified. That means we have
    $r_{\rm PA}=1$, not 0. This deserves a clarification.
    $r_{\rm PA}=\lim_{m\to\infty}\frac{k(m)}{m}=1$ states
    that the secret key is equivalent to the raw key in the infinite
    limit for both Alice and Bob and Eve, what corresponds to
    $I_{AB}=I_{AE}(D_{Max\rm -CM})=1$, for $D_{Max\rm -CM}=0.5$. So,
    $k_{\rm PA}(m)$ should not be used as a secret key, but that does
    not mean that we can infer $k_{\rm PA}(m)=0$. After {\rm PA} both
    parties have the same $r_{\rm PA}=1$ and discarding $k_{\rm PA}(m)$
    does not turn $r_{\rm PA}$ to zero. Discarding the key is based
    on Alice and Bob's estimation from the {\rm CM}, i.e., from
    outside of the {\rm MM} space of calculation. The way of
    calculating $k_{\rm PA}(m)$ so as to include discarding of
    estimated bits both parties might possess should follow from an
    adequately elaborated {\rm PA} procedure and its algorithms.
    A starting step should be a predefined $D_{Max\rm -CM}<0.5$ and
    its inclusion in the protocol via $I_{MaxAE}=I_{AE}(D_{Max\rm -CM})$.
    That would give us a {\em conditional security\/} of the
    protocol.}
\end{itemize}

Taken together, the analysis carried out in \cite{lu-cai-11} is
applicable and correct for the attacks on two-way protocols in which
Eve reads off the states of qubits with the help of ancillas but is
inapplicable to MITM attacks. Therefore, their Proof is not
universal, i.e., cannot be considered {\em unconditional}.

Can Alice and Bob still achieve reliable security of their
two-way protocol? To answer this let us first compare one-way
(e.g., BB8) and two-way protocols. 

Under standard attacks on one way protocols, Eve is left with less
and less information about the key when she approaches the critical
disturbance $D_{crit\rm{-MM}}=0.11$, i.e., the messages she snatches
end up scrambled, up to 50\% of the time. But Eve also scrambles Bob
and Alice's messages so that after PA, half of the messages are
deleted and half coincide. So, neither party is left with any
usable bit. 

In two-way protocols under MITM it is different. Eve does not
scramble Bob and Alice's messages at all at higher and higher
values of $D_{\rm CM}$ and the longer she is in the line
the more messages she copies; for $D_{Max\rm -CM}=0.5$ their
secret keys are identical and no bare {\rm PA} (hashing only)
can change that.  

However, with two-way protocols, when Eve is not in the line
all the time, Alice and Bob carry out the PA, guided by
the level of disturbance, i.e., the error rate in the CM\@. The
standard PA uses a binary string of obtained messages to produce a
new shorter strings via universal hashing. Alice randomly chooses
a (permutation) function $f:\{0,1\}^m\to\{0,1\}^{k(m)}$  among
some universal$_2$ class of functions. She then sends both $f(x)$
and a description of $f$ to Bob via universal hashing. After
computing $f(y)$, where $y$ is his corresponding string, Bob
checks whether it agrees with $f(x)$. If it does, a basic
property of universal hashing allows them to assume that $x=y$
\cite[p.~214]{bennett-uncond-sec-ieee-95}. 

The problem emerges with this version of the PA, i.e., with
algorithms it makes use of, because Alice and Bob should be
able to estimate the length $k(m)$ of the secure key with
respect to the length of the raw key $m$ which would guarantee
them that Eve is not in possession of a single bit of $k(m)$,
but via their bare (``blind'') PA they always get $x=y$, i.e.,
they do not have a benchmark for estimating the amount of bits
Eve lost. The PA procedures elaborated in the literature do not
help since they are made for one-way protocols (BB84, B92, etc.)
and are rather involved and
intricate---Cf.~\cite{renner-koening-05}. We have
not found a PA procedure elaborated for two-way protocols in the
literature. And what makes it challenging is the asymptotic
approach of $I_{AE}(D_{\rm CM})$ to $I_{AB}=1$ shown in
Fig.~\ref{fig:dist}(b), which is absent in the BB84 protocol---see
Fig.~\ref{fig:dist}(a), on the one hand, and the high $D$, on the
other. Whether we can find an efficient PA algorithm for two-way
protocols remains to be seen. 

A special kind of an aBB84-like protocol in which the $\mathbb X$
basis serves as MM and $\mathbb Z$ as CM proposed by
Bunandar et al. \cite{bunandar-18} can help us to better understand
the problem of unlimited $D$. We call the protocol a
message-control-(a)symmetric BB84 (mcasBB84) protocol.
In Table~\ref{T:comp} we compare the properties of the BB84,
two-way, and mcasBB84 protocols under MITM attacks.

\begin{table*}[ht]
\center
\setlength{\tabcolsep}{3pt}
\renewcommand{\arraystretch}{1.25}
\begin{tabular}{|c|c|c|c|c|}
\hline
&BB84&pp&LM05&mcasBB84-MITM
\\
\hline 
mode(s)
&MM
&\minibox{MM + CM}
&\minibox{MM + CM}
&\minibox{MM + CM}
  \\
  \rule{0pt}{17pt}
disturbance
&$0\le D_{\rm MM}\le 0.5$
&\minibox{\hfil$D_{\rm MM}=0$\\ $0\le D_{\rm CM}\le 0.5$}
&\minibox{\hfil$D_{\rm MM}=0$\\ $0\le D_{\rm CM}\le 0.5$}
&\minibox{\hfil$D_{\rm MM}=0$\\ $0\le D_{\rm CM}\le D_{pd{\rm -CM}}$}
\\
\minibox{maximal\\ disturbance}
&$D_{critical{\rm -MM}}=0.11$
&?
&?
&$D_{pd{\rm -CM}}$
  \\
\rule{0pt}{13pt}  
secure
&for $D_{\rm MM}<0.11$
&for $D_{\rm CM}<\ ?$
&for $D_{\rm CM}<\ ?$
&for $D_{\rm CM}<D_{pd{\rm -CM}}$
\\
\rule{0pt}{20pt}
\minibox{mutual\\ information}
&\minibox{$I_{AB}(D_{\rm MM})$, $I_{AE}(D_{\rm MM})$}
&\minibox{\hfil$I_{AB}=1$\\ $0\le I_{AE}(D_{\rm CM})<1$}
&\minibox{\hfil$I_{AB}=1$\\ $0\le I_{AE}(D_{\rm CM})<1$}
&\minibox{\hfil$I_{AB}=1$\\ $0\le I_{AE}(D_{\rm CM})<I_{AE}(D_{pd{\rm -CM}})$}
\\
\minibox{photon\\ distance}
&$L$
&4$L$
&2$L$
&$L$
\\
\minibox{trans-\\ mittance}
&$\cal T$
&${\cal T}^4$
&${\cal T}^2$
&$\cal T$
\\
\hline
\end{tabular}
\renewcommand{\arraystretch}{1}
\caption{Properties of a symmetric BB84-like protocol under an
  arbitrary attack compared with properties of pp, LM05, and
  asymmetric mcasBB84 protocols under MITM. For the pp and LM05
  protocols $D<0.5$ means that Eve is in the line only a portion
  of the time and $D=0.5$ that she is in the line all the time.
  $D_{pd\rm -CM}$ is a predetermined threshold value of $D<0.5$
  for the mcasBB84-MITM \cite{bunandar-18} protocol.}  
\label{T:comp}
\end{table*}

Let us consider a MITM attack on mcasBB84 (mcasBB84-MITM) which
Eve carries out so as to measure and resend all qubits in the
$\mathbb X$ basis. Since Eve receives only $|0\rangle$ and
$|1\rangle$ messages and resends them unchanged, she does not
introduce any disturbance in the MM and $I_{AB}=1$. Eve's $I_{AE}$
rises with her increased presence in the line. If she were in the
line all the time, we would have $D_{\rm CM}=0.5$ and $I_{AE}=1$.
But the protocol does not allow that. Instead, it predetermines a
threshold value $D_{pd\rm CM}$ and if $D_{\rm CM}>D_{pd\rm CM}$ Alice
and Bob will abort it as specified by Bunandar et
al.~\cite[p.~7]{bunandar-18}. The protocol is an
adaptation of the three-state aBB84 protocol which makes use of
both $\mathbb X$  and $\mathbb Z$ bases for MM as in the standard
BB84 only with two additional decoy settings put forward by Lim,
Curty, Walenta, Xu and Zbinden \cite{lim-14} which itself builds
on other decoy-state pioneering methods as, e.g., the one proposed
by Wang \cite{wang-05a}. The idea of a predetermined threshold value
$D_{pd\rm CM}$ is taken over from \cite{mizutani-15}. $D_{pd\rm CM}$
serves \cite{bunandar-18} to calculate a conditional security.
The calculations determines which maximal $D_{pd\rm CM}$ is
acceptable for an implementation. Apart from solving
two-way-protocol maximal $D$ problem, the mcasBB84-MITM has
another big advantage (with respect to exponential attenuation of
photons in optical fibres) that its photons cover the same distance
as the original BB84 ($L$), i.e., half the distance LM05 photons
cover (2$L$) and a quarter of the distance pp photons cover (4$L$).

\bigskip
\section{Conclusion}
\label{sec:conclusion}

To summarise, we considered man-in-middle (MITM) attacks on two
kinds of two-way QKD protocols (pp with entangled photons and LM05
with single photons) in which an undetectable Eve can decode all
the messages in the message mode (MM) and showed that the mutual 
information between Alice and Bob is not a function of disturbance
in the MM, since there is no disturbance in the MM, but is equal to
unity no matter whether Eve is in the line or not. Eve induces a
disturbance ($D_{\rm CM}$) only in the control mode (CM). In a way,
Alice's sending of the key is equivalent to sending an unencrypted
plain text (via photons obtained by and returned to Bob) secured
by an indicator of Eve's presence. That burdens the protocols with
the following drawbacks under a MITM attack: 
\begin{itemize}
\item the photons must cover the double distance than in an
  equivalent one-way BB84-like protocol (mcasBB84) which also has
  analogous MM and CM modes;
\item while the BB84 protocol is unconditionally secure, and its
  revised version, the macasBB84 protocol proposed recently is
  only conditionally secure, the proof of unconditional security
  of the LM05 protocol given in \cite{lu-cai-11} is not valid, as
  shown in details in Sec.~\ref{sec:security}; the mcasBB84
  protocol has a predetermined threshold value of the CM
  disturbance ($D_{pdCM}$) at which Bob and Alice must abort the
  protocol whilst the considered two-way protocols do not have
  such a critical CM disturbance at which to abort the protocol;
\item since there are no errors in the MM mode, the privacy
  amplification (PA) is the only way to establish the security of
  the protocols and again in contrast to the mcasBB84 no PA
  procedures for the two-way protocols have been provided in the
  literature;
\end{itemize}

Let us elaborate on these point in the reverse order. 

In the two-way protocols the mutual information between Alice and
Bob is always greater than or equal to the one between Alice and
Eve. When they are equal, i.e., when Eve is in the line all the
time, then Alice and Bob and Eve have identical messages and there
is no privacy which can be amplified and PA cannot erase key bits
Eve has snatched. For a $D_{\rm CM}<0.5$, but close to 0.5, Alice and
Bob do not have a procedure and algorithms to obtain the secret key
of which Eve possess almost all bits. Note that $I_{AE}(D_{\rm CM})$
approaches $I_{AE}=1$ asymptotically and that a maximal $D_{CM}$
after which Alice and Bob have to abort the protocol is not defined.

This is related to our analysis (in Sec.~\ref{sec:security})
of the security proof given in \cite{lu-cai-11} which the authors
call unconditional. In the analysis in Sec.~\ref{sec:security} we
show that their proof does not cover the man-in-the-middle attack
(MITM) and that therefore cannot be called ``unconditional.'' 

To better understand the problem of absence of a maximal
tolerable $D_{\rm CM}$ (after which Alice and Bob have to abort
the protocol) in two-way protocols, in Sec.~\ref{sec:security}
we compare protocol with a newly proposed one-way asymmetric
BB84-like protocol \cite{bunandar-18} (mcasBB84) in which
$\mathbb X$ basis serves as MM and $\mathbb Z$ basis as CM under
a MITM attack (mcasBB84-MITM). We show that mcasBB84-MITM without
defined maximal $D_{\rm CM}$ would be completely equivalent to
two-way protocols under MITM\@. But the mcasBB84 protocol resolves
the problem of a maximal $D_{\rm CM}$ by means of a predetermined 
threshold value $D_{pd\rm CM}$. When $D_{\rm CM}>D_{pd\rm CM}$ Alice
and Bob abort the protocol \cite[p.~7]{bunandar-18}. The security
calculated for such $D_{pd\rm CM}$ \cite[pp.~7-10]{bunandar-18},
i.e., an elaborated PA procedure, can be called a ``conditional
security.'' An additional advantage (with respect to exponential
attenuation of photons in optical fibres) of mcasBB84 is that
photons do not travel from Bob to Alice and back to Bob, but only
from Alice to Bob (see Table~\ref{T:comp}). 

A similar solution for two-way protocols would be to redesign
the protocol so as to either calculate a critical $D_{crit\rm -CM}$
at which Alice and Bob would be able to erase all bits Eve might
have possessed via privacy amplification (PA) or to predetermine
threshold value of the disturbance in the CM, $D_{pd\rm -CM}$, for
which PA calculations might be carried out. The former calculation,
if possible, would provide us with an unconditionally security and
the latter one would provide us with a conditional security of
the protocols. How to do either of the calculations is an open
question, but we conjecture that the former calculation is not
feasible. 

\begin{acknowledgments}
Supported by the Ministry of Science and Education
of Croatia through the Center of Excellence for Advanced
Materials and Sensing Devices (CEMS) funding, and by
MSE grants Nos. KK.01.1.1.01.0001 and 533-19-15-0022.
Computational support was provided by the cluster Isabella of
the Zagreb University Computing Centre and by the Croatian
National Grid Infrastructure (CRO-NGI). Financial supports by
the Alexander von Humboldt Foundation as well as by the German
Research Foundation (DFG) and the Open Access Publication Fund
of Humboldt-Universität zu Berlin are acknowledged.
\end{acknowledgments}

\bigskip 


\end{document}